\documentclass[12]{article}
\usepackage{fleqn}
\usepackage{graphicx}
\begin{document}
\Large
\begin{center}{\bf
Multi-Line Geometry of Qubit-Qutrit and Higher-Order\\ Pauli
Operators}
\end{center}
\vspace*{-.0cm}
\begin{center}
Michel Planat$^{\dag }$, Anne-C\'{e}line Baboin$^{\dag }$
 and Metod Saniga$^{\ddag}$
\end{center}
\vspace*{-.3cm} \normalsize
\begin{center}
\vspace*{.1cm} $^{\dag}$Institut FEMTO-ST, CNRS, D\' epartement LPMO, 32 Avenue de
l'Observatoire\\ F-25044 Besan\c con, France\\
(michel.planat@femto-st.fr)

\vspace*{.1cm}
 and

$^{\ddag}$Astronomical Institute, Slovak Academy of Sciences\\
SK-05960 Tatransk\' a Lomnica, Slovak Republic\\
(msaniga@astro.sk)

\end{center}

\vspace*{-.1cm} \noindent \hrulefill

\vspace*{.1cm} \noindent {\bf Abstract.} The commutation relations
of the generalized Pauli operators of a qubit-qutrit system are
discussed in the newly established graph-theoretic  and
finite-geometrical settings. The dual of the Pauli graph of this
system is found to be isomorphic to the projective line over the
product ring $\mathcal{Z}_2\times\mathcal{Z}_3$. A ``peculiar"
feature in comparison with two-qubits is that two distinct
points/operators can be joined by more than one line. The
multi-line property is shown to be also present in the
graphs/geometries characterizing two-qutrit and three-qubit Pauli
operators' space and surmised to be exhibited by any other
higher-level quantum system.
\\

\noindent
{\bf PACS Numbers:} 03.65.-w, 03.65.Fd, 02.10.Ox, 02.40.Dr\\
{\bf Keywords:} Generalized Pauli Operators --- Pauli Graphs
--- Finite Projective Geometries

\noindent
\hrulefill
\section{Introduction}

Although being central to topics such as the derivation of
complete sets of mutually unbiased bases \cite{Bandyo,PlanatMUBs},
or to an in depth understanding of quantum entanglement
\cite{Klimov,sigma}, the commutation relations between the
generalized Pauli operators of finite-dimensional quantum systems
are still not well understood. Recently, considerable progress has
been made in this respect by employing finite geometries such as
finite projective lines \cite{sigma,ThMathPh}, generalized
quadrangles \cite{Saniga0,Pauligraphs,Payne} and polar spaces
\cite{Saniga1,KThas} to treat dimensions $d=2^N$ and, most
recently \cite{planatPrague,Pauligraphs}, also the case of $d=9$. In this
paper, after introducing the basic notation about generalized
Pauli operators and Pauli graphs and brief recalling the
established results, we will first have a look at the smallest
composite dimension, $d=6$, as this is the first case where we
expect to find serious departures from what is known about Hilbert
spaces whose dimension is a power of a prime. We shall, indeed,
find that the finite geometry here is more intricate, exhibiting
more than one line sharing two distinct points. In light of this
finding,  we then revisit the $d=3^2$ case and, finally, briefly
address the case of $d=2^3$.

A complete orthonormal set of operators in a $p$-dimensional
Hilbert space ($p$ a prime number) is of cardinality of $p^2-1$.
These operators can be derived from the {\it shift} and {\it
clock} operators $X$ and $Z$
\begin{equation}
X|n\rangle =|n+1\rangle,~~Z|n\rangle =\omega_p^n|n\rangle~~\mbox{with}~~\omega_p=\exp{(2i\pi/p)},
\label{ShiftandClock1}
\end{equation}
as follows \cite{Bandyo,Klimov},
\begin{eqnarray}
&\{Z^k\},~~k=1,...,p-1,\nonumber \\
&\{(XZ^m)^k\},~~k=1,...,p-1,~~m=0,...,p-1.
\label{ShiftandClock2}
\end{eqnarray}
and grouped together into $p+1$ disjoint classes, each comprising
$p-1$ pairwise commuting members. The common eigenstates of
distinct classes form different sets of mutually unbiased bases
\cite{Bandyo, Klimov}. As a result, such sets of mutually unbiased
bases are complete for Hilbert spaces of the corresponding
dimensions.

\noindent The simplest ($p=2$) case corresponds to qubits. The
orthonormal set comprises the standard Pauli matrices $\sigma_i=
(I_2,\sigma_x, \sigma_y,\sigma_z)$, $i \in \{1,2,3,4\}$, where
$I_2=\left(\begin{array}{cc}1 & 0 \\0 & 1\\\end{array}\right)$,
$\sigma_x=\left(\begin{array}{cc}0 & 1 \\1 &
0\\\end{array}\right)$, $\sigma_z=\left(\begin{array}{cc}1 & 0 \\0
& -1\\\end{array}\right)$ and $\sigma_y=i \sigma_x \sigma_z$. In
the next case, $p=3$, one gets
$\sigma_j=\{I_3,Z,X,Y,V,Z^2,X^2,Y^2,V^2\},~~j \in \{1,\dots,9\}$,
where $I_3$ is the $3 \times 3$ unit matrix,
$Z=\left(\begin{array}{ccc}1 & 0&0
\\0 & \omega&0\\0&0& \omega^2\\\end{array}\right)$,
$X=\left(\begin{array}{ccc}0 & 0&1 \\1 & 0&0\\0&1&
0\\\end{array}\right)$, $Y=XZ$, $V=X Z^2$ and $\omega=\exp\left(2
i \pi/3\right)$. The implementation of this procedure for an
arbitrary prime $p$ is then straightforward. Going to Hilbert
spaces of prime-power dimensions $d=p^N$, $N \geq 2$, the
$(d^2-1)$ generalized Pauli operators are similarly partitioned
into $(d+1)$ disjoint sets, each one composed of a maximum set of
$(d-1)$ mutually commuting members.

In our graph-geometrical approach
\cite{sigma}--\cite{Pauligraphs}, \cite{Saniga1}, the generalized
Pauli operators are identified with the points/vertices and
maximum mutually commuting members of them with the edges/lines (or
subspaces of higher dimensions) of a specific Pauli graph/finite
incidence geometry so that the structure of the operators' space
can fully be inferred from the properties of the Pauli
graph/finite geometry in question. It has been found that the
operators' space characterizing two-qubits is isomorphic to the
generalized quadrangle of order two \cite{Saniga0,Pauligraphs} and
that $N$-qubits ($N > 2$) are intimately related with symplectic
polar spaces of rank $N$ and order two \cite{Saniga1}. A crucial
role in this discovery turned out to be the concept of projective
lines defined over rings \cite{Saniga2}, as the generalized
quadrangle of order two is embedded as a sub-geometry in the
distinguished projective line defined over the full two-by-two
matrix ring with coefficients in $\mathcal{Z}_2$ \cite{Saniga0}.
When analyzing in a similar fashion two-qutrit ($d=3^2$) case
\cite{Pauligraphs,planatPrague}, it turned out to be convenient to pass to the
{\it dual} of the Pauli graph, i.e., to the graph whose vertices
are represented by the {\it maximum commuting subsets} of the
operators, two vertices being adjacent if the corresponding
maximum sets have an operator in common. This move is also a
fruitful and necessary one when addressing properly the simplest
composite ($d=6$) case, which is the subject of the next section.

\begin{figure}[h]
\centerline{\includegraphics[width=10.4truecm,clip=]{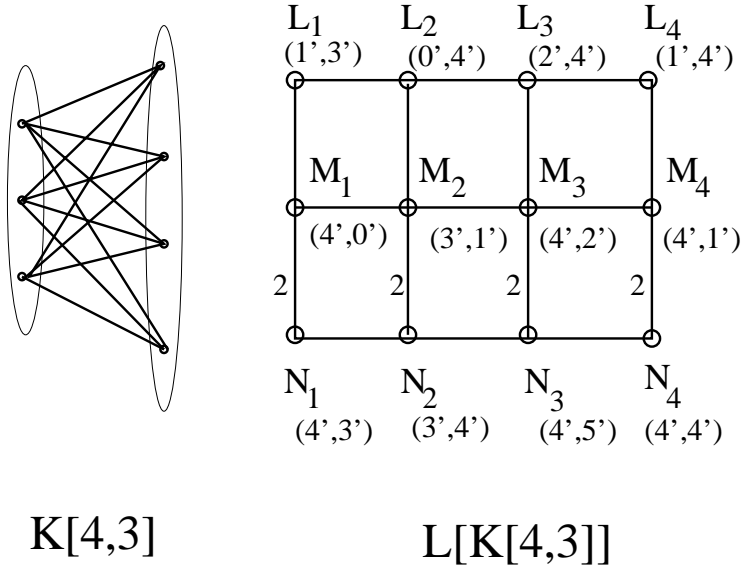}}
\caption{The graph $\mathcal{W}_6$, the dual of the Pauli graph
$\mathcal{P}_6$, as the line graph of the complete bipartite graph
$K[4,3]$. The mutually unbiased bases correspond to the lines of
the Pauli graph $\mathcal{P}_6$ which are not concurrent, i.e., to
the vertices of $\mathcal{W}_6$ which are not adjacent. Lines
$L_i$ (as well as $M_i$ and $N_i$) mutually intersect at a single
point, whereas $L_i$ and $M_i$ (as well as $L_i$ and $N_i$ and
$M_i$ and $N_i$) have two points in common; this means that the
adjacency in $\mathcal{W}_6$ is of two different ``weights" (``1"
and ``2", the latter explicitly indicated). This $3 \times 4$ grid
can also be regarded as the projective line over the ring
$\mathcal{Z}_2 \times \mathcal{Z}_3$. } \label{grid}
\end{figure}
\begin{figure}[h]
\centerline{\includegraphics[width=5.0truecm,clip=]{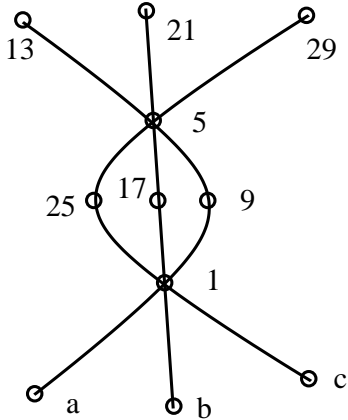}}
\caption{A schematic sketch of the point-set structure of
$\mathcal{S}_1=\{L_1,M_1,N_1\}$. } \label{multi}
\end{figure}

\vspace*{.1cm}

\section{The ``multi-line" geometry of a qubit-qutrit}

For the sextic ($d=6$) systems, one has $6^2-1=35$ generalized
Pauli operators\footnote{It is obvious that the geometry of a
qutrit-qubit system is isomorphic to that of the qubit-qutrit
case.}
\begin{equation}
\Sigma_6^{(i,j)}=\sigma_i \otimes \sigma_j,~~i\in
\{1,\ldots,4\},~~j\in \{1,\ldots, 9\},~~(i,j)\neq(1,1),
\end{equation}
which can be conveniently labelled as follows: $1=I_2 \otimes
\sigma_1$, $2=I_2 \otimes \sigma_2$, $\cdots$, $8=I_2 \otimes
\sigma_8$ , $a=\sigma_z \otimes I_2$, $9=\sigma_z \otimes
\sigma_1$,\ldots, $b=\sigma_x \otimes I_2$, $17=\sigma_x \otimes
\sigma_1$,\ldots , $c=\sigma_y \otimes I_2$,\ldots$,32=\sigma_y
\otimes \sigma_8$. Joining two distinct mutually commuting
operators by an edge, one obtains the corresponding Pauli graph
$\mathcal{P}_6$. It is straightforward to derive twelve maximum
commuting sets of operators,
\begin{eqnarray}
&L_1=\{1,5,a,9,13\},~~L_2=\{2,6,a,10,14\},~~L_3=\{3,7,a,11,15\},~~L_4=\{4,8,a,12,16\}, \nonumber \\
&M_1=\{1,5,b,17,21\},~~M_2=\{2,6,b,18,22\},~~M_3=\{3,7,b,19,23\},~~M_4=\{4,8,b,19,24\}, \nonumber\\
&N_1=\{1,5,c,25,29\},~~N_2=\{2,6,c,26,30\},~~N_3=\{3,7,c,27,31\},~~N_4=\{4,8,c,28,32\}\nonumber,
\end{eqnarray}
which are regarded as lines of the associated finite geometry.
Then, considering these lines as the vertices of the dual graph,
$\mathcal{W}_6$, with an edge joining two vertices representing
concurrent lines, we arrive at a grid-like graph shown in
Fig.~\ref{grid}, right. This graph corresponds to $L[K(4,3)]$,
i.\,e., to the line graph of the bipartite graph $K(4,3)$; it is a
regular graph with spectrum $\{-2^6,1^3,2^2,5\}$. Mutually
unbiased bases correspond to mutually disjoint lines and, hence,
non-adjacent vertices of $\mathcal{W}_6$;  from Fig.~\ref{grid},
right, it is readily seen that a maximum of three of them arise,
as expected \cite{Grassl}. It is also worth mentioning that
$\mathcal{W}_6$ can be regarded the projective line over the
product ring $\mathcal{Z}_2 \times \mathcal{Z}_3 \cong Z_6$, where
the term ``adjacent" means ``neighbor" (see Appendix for more
details).

Returning back to $\mathcal{P}_6$ and we can show that the
associated geometry resembles to some extent that of a finite
$(0,1)$-geometry , i.\,e., the point-line incidence structure
where (i) two distinct points are contained in at most one line
and where (ii) given a line and a point not on the line ({\it aka}
an anti-flag), there exists either zero or one line through the point
that intersect the line in question \cite{Payne,Batten}. For
although we saw that our geometry is endowed with ``multi-lines"
(i.\,e., lines sharing more than one point) and so violates the
first axiom of a $(0,1)$-geometry , we still find that the
connection number for each anti-flag is either zero or one. Hence, we can define
an analogue of a geometric hyperplane (see, e.\,g., \cite{Saniga3}) as a subset of points of our
$\mathcal{P}_6$ geometry such that whenever its two points lie on
a line then the entire line lies in the subset. Then we readily
verify that the sets of points located on the following four
triples of lines
\begin{equation}
\mathcal{S}_i=\{L_i,M_i,N_i\},~~(i=1,2,3,4)
\end{equation}
represent each a ``geometric hyperplane" of our geometry, as
illustrated in Fig.~\ref{multi} for $\mathcal{S}_1$. The three
lines of $\mathcal{S}_i$ intersects at two points and the
connection number for any of its anti-flags is one.

\begin{figure}[h]
\centerline{\includegraphics[width=9.0truecm,clip=]{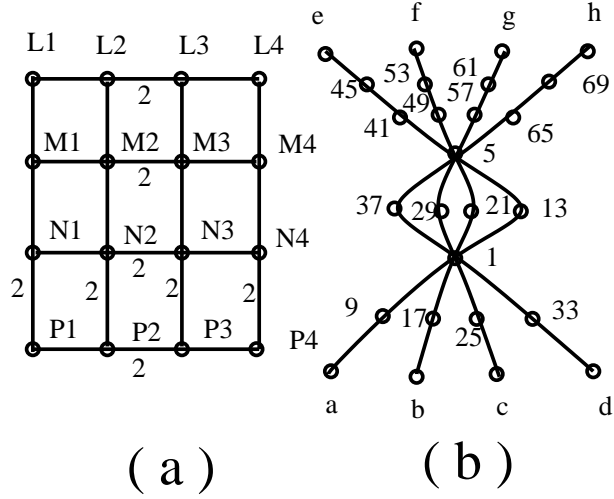}}
\caption{(a) An example of a grid in $\mathcal{W}_9$; each edge is
of the same weight as every pair of concurrent lines intersect at
two points as indicated. (b) A set of four lines
${L_i,M_i,N_i,P_i}$ form a ``multi-line" subset of the geometry
associated with $\mathcal{P}_9$, here illustrated for $i=1$.}
\label{multi1}
\end{figure}
\begin{figure}[h]
\centerline{\includegraphics[width=9.4truecm,clip=]{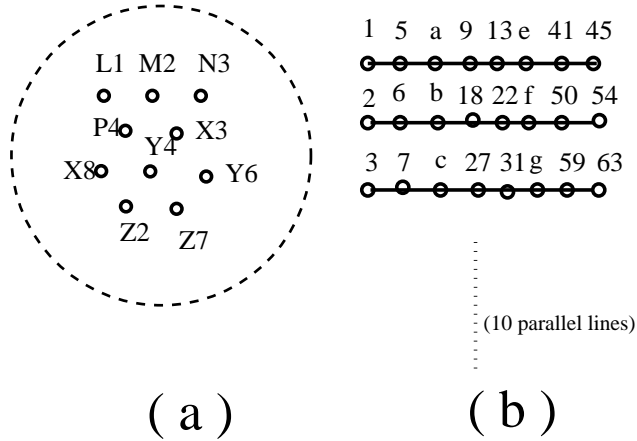}}
\caption{(a) An ovoid of $\mathcal{W}_9$, i.\,e., a set of ten
pairwise non-collinear points and (b) its image in the geometry of
$\mathcal{P}_9$, i.\,e., a set of ten pairwise disjoint (or
parallel) lines.} \label{multi2}
\end{figure}

\section{The ``multi-line" geometry behind two-qutrits}

For the two-qutrit system ($d=9$)  one has the $9^2-1=80$
generalized Pauli operators
\begin{equation}
\Sigma_9^{(i,j)}=\sigma_i \otimes \sigma_j,~~i\in \{1,\ldots,
9\},~~j\in \{1,\ldots, 9\}~~\mbox{and}~~(i,j)\neq(1,1)
\end{equation}
which can be labelled as follows: $1=I_2 \otimes \sigma_1$, $2=I_2
\otimes \sigma_2$, $\cdots$, $8=I_2 \otimes \sigma_8$, $a=\sigma_1
\otimes I_2$, $9=\sigma_1 \otimes \sigma_1$,\ldots, $b=\sigma_2
\otimes I_2$, $17=\sigma_2 \otimes \sigma_1$,\ldots , $c=\sigma_3
\otimes I_2$,$\ldots$, $h=\sigma_8 \otimes I_2$,$\ldots$,
$72=\sigma_8 \otimes \sigma_8$. The associated Pauli graph
$\mathcal{P}_9$ is regular, of degree $25$, and its spectrum is
$\{-7^{15},-1^{40},5^{24},25\}$ \cite{planatPrague}. It is a
straightforward, but rather cumbersome, task to derive the
following 40 maximum commuting sets of operators

\footnotesize
\begin{eqnarray}
&L_1=\{1,5,a,9,13,e,41,45\},~~L_2=\{2,6,a,10,14,e,42,46\},~~L_3=\{3,7,a,11,15,e,43,47\},\nonumber \\
&L_4=\{4,8,a,12,16,e,44,48\},~~M_1=\{1,5,b,17,21,f,49,53\},~~M_2=\{2,6,b,18,22,f,50,54\},\nonumber\\
&M_3=\{3,7,b,19,23,f,51,55\},~~M_4=\{4,8,b,20,24,f,52,56\},~~N_1=\{1,5,c,25,29,g,57,61\},\nonumber\\
&N_2=\{2,6,c,26,30,g,58,62\},~~N_3=\{3,7,c,27,31,g,59,63\},~~N_4=\{4,8,c,28,32,g,60,64\},\nonumber \\
&P_1=\{1,5,d,33,37,h,65,69\},~~P_2=\{2,6,d,34,38,h,66,70\},~~P_3=\{3,7,d,35,39,h,67,71\},\nonumber\\
&P_4=\{4,8,d,36,40,h,68,72\},\nonumber\\
&X_1=\{9,22,32,39,45,50,60,67\},~~X_2=\{10,17,27,40,46,53,63,68\},~~X_3=\{11,20,30,33,47,56,58,69\},\nonumber\\
&X_4=\{12,23,25,34,48,51,61,70\},~~X_5=\{13,18,28,35,41,54,64,71\},~~X_6=\{14,21,31,36,42,49,59,72\},\nonumber\\
&X_7=\{15,24,26,37,43,52,62,65\},~~X_8=\{16,19,29,38,44,55,57,66\},\nonumber\\
&Y_1=\{9,23,30,40,45,51,58,68\},~~Y_2=\{10,19,32,33,46,55,60,69\},~~Y_3=\{11,22,25,36,47,50,61,72\},\nonumber\\
&Y_4=\{12,17,26,39,48,53,62,67\},~~Y_5=\{13,20,27,34,41,56,63,70\},~~Y_6=\{14,23,28,37,42,51,64,65\},\nonumber\\
&Y_7=\{15,18,29,40,43,54,57,68\},~~Y_8=\{16,21,30,35,44,49,58,71\},\nonumber\\
&Z_1=\{9,24,31,38,45,52,59,66\},~~Z_2=\{10,24,25,35,46,52,61,71\},~~Z_3=\{11,17,28,38,47,53,64,66\},\nonumber\\
&Z_4=\{12,18,31,33,48,54,59,69\},~~Z_5=\{13,19,26,36,41,55,62,72\},~~Z_6=\{14,20,29,39,42,56,57,67\},\nonumber\\
&Z_7=\{15,21,32,34,43,49,60,70\},~~Z_8=\{16,22,27,37,44,50,63,65\}.
\end{eqnarray}
\normalsize \noindent From there we find that the dual graph
$\mathcal{W}_9$ consists of 40 vertices and has spectrum
$\{-4^{15},2^{24},12\}$, which are the characteristics identical
with those of the generalized quadrangle of order three formed by
the points and lines of a parabolic quadric $Q(4,3)$ in $PG(4,3)
$\cite{Payne}. The quadrangle $Q(4,3)$, like its two-qubit
counterpart, exhibits all the three kinds of geometric
hyperplanes, a grid (of order (3,1)), an ovoid, and a perp-set
(see, e.\,g., \cite{Saniga3}), and these three kinds of subsets
are all indeed found to sit inside the finite geometry associated
with $\mathcal{W}_9$ \cite{planatPrague}.

\noindent A grid of $\mathcal{W}_9$ is shown in Fig.~3a. The
subsets of points lying on the following quadruples of lines
\begin{equation}
\mathcal{S}_i=\{L_i,M_i,N_i,P_i\},~~i=1, 2, 3, 4,
\end{equation}
then represent ``multi-line hyperplanes" in the geometry
associated with $\mathcal{P}_9$, as depicted in Fig.~3b; the
``multi-line hyperplane" corresponding to the grid as a whole is
obtained by taking all the four copies $S_i$, having the eight
reference points $a$, $b$,\ldots, $h$ in common. Fig.~4 shows an
ovoid of $\mathcal{W}_9$ (a) and its counterpart in the
$\mathcal{P}_9$ geometry (b); the bases associated with the
operators on any two distinct lines of this set are mutually
unbiased and, when taken together, they thus form a complete set
for this dimension. Finally, Fig.~5 gives an example of a perp-set
of $\mathcal{W}_9$ (a) together with the detailed structure of its
``multi-line" counterpart in $\mathcal{P}_9$ (b).

\begin{figure}[pth!]
\centerline{\includegraphics[width=10.0truecm,clip=]{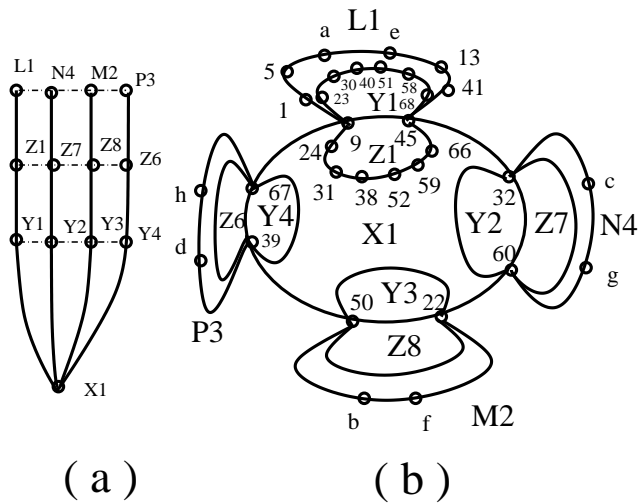}}
\caption{(a) One of the perp-sets in $\mathcal{W}_9$ and (b) its
dual ``multi-line" twin; in the latter case, the structure of only
one of the four ``multi-line" subsets is shown in full detail.}
\label{multi3}
\end{figure}

\section{Hints about ``multi-line" configurations pertinent to three-qubits}
It is worth observing that whereas in the case of two-qubits there
exists a perfect duality between the operators (points) and
maximum commuting (sub)sets of them (lines)
\cite{Saniga0,Pauligraphs}, this property is lost when we pass to
higher level quantum systems; thus, as we have seen, in the
qubit-qutrit case we have 35 operators but only 12 lines, whereas
two-qutrits give 80 operators and 40 lines. We surmise that it is
this loss of duality which enables the emergence of ``multi-line"
objects in the corresponding finite geometries.

To partly justify this surmise, we will briefly address the case
of three-qubits. The $4^3-1=63$ tensor products of the classical
Pauli matrices $\sigma_i \otimes \sigma_j \otimes \sigma_k$,
$[i,j,k=1,2,3,4$, $(i,j,k)\neq (1,1,1)]$ form the vertices and
their commuting pairs the edges of the strongly regular graph,
$\mathcal{P}_8$, of degree $30$ and spectrum
$\{-5^{27},3^{35},30\}$. Employing the same strategy for labelling
the operators as in the preceding sections, i.\,e., $1=I_2 \otimes
I_2 \otimes \sigma_1$, $2=I_2 \otimes I_2 \otimes\sigma_2$,\ldots,
$15=I_2 \otimes \sigma_3 \otimes \sigma_3$, $a=\sigma_1\otimes I_2
\otimes I_2$, etc., one finds out that the lines of the associated
geometry, of cardinality seven each, allow indeed the existence
``multi-line hyperplanes." A portion of one of them is shown in
Fig.\,6, fully in the $\mathcal{W}_8$ (a) and partially in the
$\mathcal{P}_8$ (b) representation; the $3 \times 3$ grid can be regarded
as a dual analogue of the classical Mermin square in the space of observables
of two-qubits, which is a crucial element in the proof of the Kochen-Specker theorem
in dimension four \cite{Mermin}.

\begin{figure}[h]
\centerline{\includegraphics[width=9.0truecm,clip=]{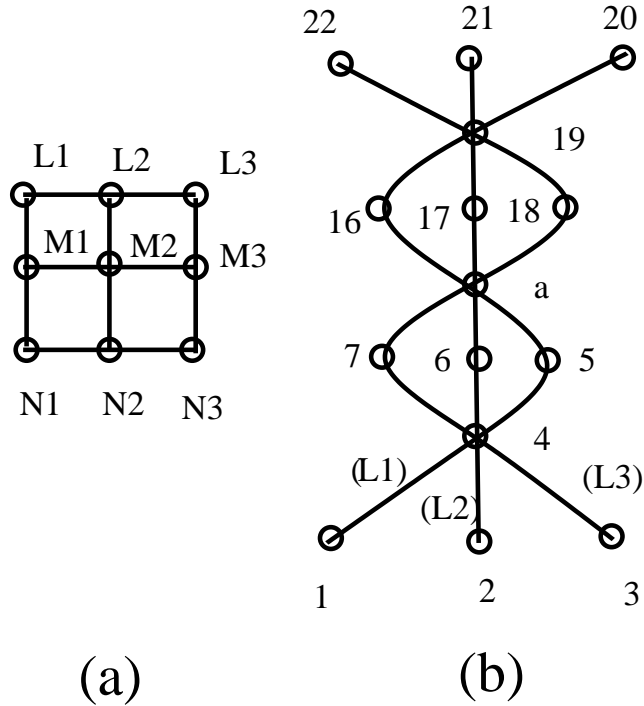}}
\caption{(a) A $3 \times 3$ grid in the $\mathcal{W}_8$
representation of the three-qubit system with adjacency of weight
three and (b) its generic part unveiled in the $\mathcal{P}_8$
perspective. Seven observables are shared by the triple of lines
in a row of the square, but only three by the triple of lines in a
column.} \label{multi3}
\end{figure}

\section{Conclusion}
Given a finite-dimensional quantum system, the set of
corresponding generalized Pauli operators and the set of maximum
commuting subsets of them can be viewed as a point-line incidence
geometry in a dual way; either regarding the operators as the
points and the maximum commuting subsets as the lines, or vice
versa. In the two-qubit case, the two pictures have been found to
be isomorphic to each other thanks to the fact that the underlying
geometry, a finite generalized quadrangle of order two, is a
self-dual object \cite{Saniga0,Pauligraphs}. This self-duality,
however, seems to disappear as we go to higher-dimensional Hilbert
spaces. An intriguing symptom of this broken symmetry is that in
one of the two representations we encounter lines sharing more
than one point in common, as illustrated here in detail for the
cases of dimension six, nine and eight. We surmise that this holds
true for any finite-dimensional quantum system except for
two-qubits.

\vspace*{.5cm} \noindent {\bf Acknowledgements}\\
\normalsize This work was partially supported by the Science and
Technology Assistance Agency under the contract $\#$
APVT--51--012704, the VEGA grant agency projects $\#$ 2/6070/26
and $\#$ 7012 (all from Slovak Republic), the trans-national
ECO-NET project $\#$ 12651NJ ``Geometries Over Finite Rings and
the Properties of Mutually Unbiased Bases" (France) and by the
CNRS--SAV Project $\#$ 20246 ``Projective and Related Geometries
for Quantum Information" (France/Slovakia).

\section*{Appendix}
\label{projline}
The concept of a projective line defined over a (finite) ring \cite{Saniga2,bh} turned out of great importance
in discovering the relevance of finite projective geometries for a deeper understanding of finite-dimensional quantum systems. This started with 
recognition of the projective line over $\mathcal{Z}_2 \times \mathcal{Z}_2$ behind a Mermin
square of two-qubits \cite{ThMathPh}, and followed by realization that the geometry of two-qubits is fully embedded within the line defined over
the smallest $2 \times 2$ matrix  ring,  $\mathcal{Z}_2^{2 \times 2}$ \cite{Saniga0}. Here we shall demonstrate that $\mathcal{W}_6$ is isomorphic to the projective line over the ring $\mathcal{Z}_2 \times \mathcal{Z}_3$.

Given an associative ring $R$ with unity and $GL(2,R)$, the general linear group of invertible two-by-two matrices with entries in $R$, a pair $(\alpha,\beta)$ is called admissible over $R$ if there exist $\gamma,\delta \in R$ such that $\left(
\begin{array}{cc}
\alpha & \beta \\
\gamma & \delta \\
\end{array}
\right) \in {\rm GL}_{2}(R)$. The projective line over $R$ is
defined as the set of equivalence classes of ordered pairs
$(\varrho \alpha, \varrho \beta)$, where $\varrho$ is a unit of
$R$ and $(\alpha, \beta)$ admissible \cite{bh}. Such a
line carries two non-trivial, mutually complementary relations of
neighbor and distant. In particular, its two distinct points $X$:
$(\varrho \alpha, \varrho \beta)$ and $Y$: $(\varrho \gamma,
\varrho \delta)$ are called {\it neighbor} if $\left(
\begin{array}{cc}
\alpha & \beta \\
\gamma & \delta \\
\end{array}
\right) \notin {\rm GL}_{2}(R)$ and {\it distant} otherwise. The
structure of the line over a finite ring can be illustrated in terms of a graph whose vertices are the points of the line and edges
join any two mutually neighbor points. 

The elements of the ring $\mathcal{Z}_2 \times \mathcal{Z}_3 \cong Z_6$ can be taken in the form $0'=(0,0)$, $1'=(0,1)$, $2'=(0,2)$, $3'=(1,0)$, $4'=(1,1)$ and 5'=(1,2), where the first element in a pair belongs to $\mathcal{Z}_2$ and the second to $\mathcal{Z}_3$. Addition and multiplication
is carried out component-wise (Table 1) and one finds that the ring contains four zero-divisors ($0'$, $1'$, $2'$, $3'$) and two units ($4'$ and $5'$). 
Employing the above-given definition, it follows that the corresponding projective line is endowed with twelve points out of which
(i) eight are represented by pairs where at least one entry is a unit, (ii) two has both entries units and (iii) two have both entries zero divisors, namely
\begin{eqnarray}
&(i)~~(4',0'),~ (4',1'),~(4',2'),~(4',3'),~(0',4'),~(1',4'),~(2',4'),~(3',4')\nonumber \\
&(ii)~~(4',4'),~(4',5'),~~~~(iii)~~(1',3'),~(3',1').\nonumber 
\end{eqnarray}
It is an easy exercise to check that the graph of the line is identical to that shown in Fig.~\ref{grid}, right (see also \cite{hav}), which implies its isomorphism to $\mathcal{W}_6$.

\begin{table}[h]
\begin{center}
\begin{tabular}{||l|cccccc||}
\hline \hline
$+$ & $0'$ & $1'$ & $2'$ & $3'$ &  $4'$ & $5'$ \\
\hline
$0'$ &   $0'$ &   $1'$ &   $2'$ & $3'$ &   $4'$ & $5'$ \\
$1'$ &   $1'$ &   $2'$ &   $0'$ & $4'$ &   $5'$ & $3'$ \\
$2'$ &   $2'$ &   $0'$ &     $1'$ & $5'$ &   $3'$ & $4'$\\
$3'$ &   $3'$ &     $4'$ &   $5'$ & $0'$ &   $1'$ & $2'$\\
$4'$ &   $4'$ &   $5'$ &   $3'$ & $1'$ &   $2'$ & $0'$\\
$5'$ &   $5'$ &     $3'$ &   $4'$ & $2'$ &   $0'$ & $1'$\\
\hline \hline
\end{tabular}~~~~~~~
\begin{tabular}{||l|cccccc||}
\hline \hline
$\times$ & $0'$ & $1'$ & $2'$ & $3'$ &  $4'$ & $5'$ \\
\hline
$0'$ &   $0'$ &   $0'$ &   $0'$ & $0'$ &   $0'$ & $0'$ \\
$1'$ &   $0'$ &   $1'$ &   $2'$ & $0'$ &   $1'$ & $2'$ \\
$2'$ &   $0'$ &   $2'$ &     $1'$ & $0'$ &   $2'$ & $1'$\\
$3'$ &   $0'$ &     $0'$ &   $0'$ & $3'$ &   $3'$ & $3'$\\
$4'$ &   $0'$ &   $1'$ &   $2'$ & $3'$ &   $4'$ & $5'$\\
$5'$ &   $0'$ &     $2'$ &   $1'$ & $3'$ &   $5'$ & $4'$\\
\hline \hline\end{tabular}~.
\caption{Addition (left) and multiplication (right) in  $\mathcal{Z}_2 \otimes \mathcal{Z}_3$.}
\label{table2}
\end{center}\end{table}

\end{document}